# Quantum Hall Effect: Current Distribution and the Role of Measurement

K. Tsemekhman, V. Tsemekhman, C. Wexler, and D. J. Thouless

*Department of Physics, University of Washington, Box 351560, Seattle, WA 98195*

Based on very general arguments we show that for any geometry of the experiment, in the regime when the measurement leads to a quantized Hall conductivity, almost the entire current injected into the sample is carried by the bulk states. The crucial role of the contacts equilibrating different edge channels is emphasized.

Since the discovery of the Quantum Hall Effect (QHE) [1] the question of how the current injected into the sample is distributed there, has attracted a lot of discussion. Two pictures — 'edge channel current' and 'distributed currents' — have been proposed. Both are based on the peculiarities of how the two-dimensional electron gas is confined in the inversion layer. The edge current picture [2,3] suggests that the Hall current flows in the narrow regions along the sample boundaries, called the edge channels. These channels are formed when Landau levels that lie below the Fermi level in the bulk, intersect it at the points close to the boundaries. This picture would obviously be true if one were talking about usual electron transport, as the edge channel states are the only extended states at the Fermi energy. The Hall current, however, is a supercurrent due to the drift of electrons in crossed electric and magnetic fields, and is not necessarily carried by the states at or even close to the Fermi level. Generally, all the states below the Fermi energy can carry the Hall current. The only requirement for an occupied extended state to carry the Hall current is a non-zero wave function in the region where there is a non-zero potential gradient. Therefore, the statement that the injected current flows along the edge channels is equivalent to saying that the Hall voltage drops entirely in these regions. On the other hand, the description in terms of distributed currents suggests that the Hall voltage drops gradually across the sample [4–9]. The experiments meant to find out which of these two descriptions is correct do not provide a definite answer. Measurements of the equilibration rates between the current carrying states [10], as well as experiments on non-local conductivity [11] seem to support the idea of edge currents. However, studies of the breakdown of the QHE [12] and direct measurements of the Hall voltage in the sample [13] favor the picture of bulk currents.

Previous theoretical studies of the current distribution in the QH sample could be applied to any two-dimensional electron system with an incompressible bulk region; they were not specific about the conditions under which the quantization of the Hall conductivity is observed. At the same time, it was realized [3,14] that the equilibration between edge channels is crucial for the exact quantization. The importance of the ideal contacts has also been emphasized in this context. In this letter, we put all these ideas together, and based on very general arguments show that in the regime when the measured Hall conductivity is *exactly quantized*, the Hall current is almost entirely carried by the states *in the bulk of the sample*.

Consider first a two-dimensional electron system with a strong magnetic field and weak disorder, confined in the $x$-direction by a self-consistent potential created by the positive background, electron charge distribution, and gate voltage. Under these conditions, disorder does not localize all the states, and the states in the center of the each Landau level remain extended. In what follows, only extended states are discussed. In the Landau gauge, with vector potential $A_y = Bx$, the electrons are free in the $y$-direction and quantized in the $x$-direction. Wave functions can be written as $e^{iky}f_n(x - kl^2)$, where $l = (\hbar c/eB)^{1/2}$ is the magnetic length and $n$ enumerates the Landau levels. In the absence of the confining potential, wave functions and energies are simply those of the harmonic oscillator: $\epsilon_n = \hbar\omega_c(n + 1/2)$, where $\omega_c = eB/mc$ is the cyclotron frequency. In the presence of the confining potential, as $x_0 = kl^2$ approaches the boundary of the sample, wave functions become asymmetric around $x_0$, and the energies of the corresponding states rise [2]. In the one-electron picture, edge channels are formed at the intersection of the Landau levels with the Fermi energy. The number of these channels is thus equal to the number of bulk Landau levels below the Fermi level. When the interaction between electrons is taken into account, regions with partial filling (compressible strips) are formed at the intersection [15,16]. A simplified picture of the energy levels and occupation numbers based on the results of the self-consistent calculations in the finite-temperature Hartree approximation is shown on Fig. 1(a,b). (Details of the calculations will be reported elsewhere.)

In equilibrium, although the net current is zero, orbital current flows along the boundaries of the sample. There are two contributions to this equilibrium current which are spatially separated [17]. The first one, the diamagnetic current, is roughly proportional to the gradient of the self-consistent confining potential, and it is non-zero only in the incompressible regions. The other one is associated with the concentration gradient, which is different from zero only in the compressible strips close to the device boundary. However, the total current carried by the partially occupied near-edge region of the $n$th Landau level is always $(n+1/2)e\omega_c/2\pi$, independent of external conditions [17]. It is certainly minus this value in the corresponding strip along the opposite edge. Therefore, only incompressible regions can support non-equilibrium current. This can be easily seen from classical considerations: non-equilibrium Hall current is carried by electrons drifting in crossed electric and magnetic fields in the direction perpendicular to both of them. This drift velocity is proportional to the local electric field, which is identically zero in the compressible regions. We will further ignore the current in the compressible regions since it is not changed by the injected *non-equilibrium* current.

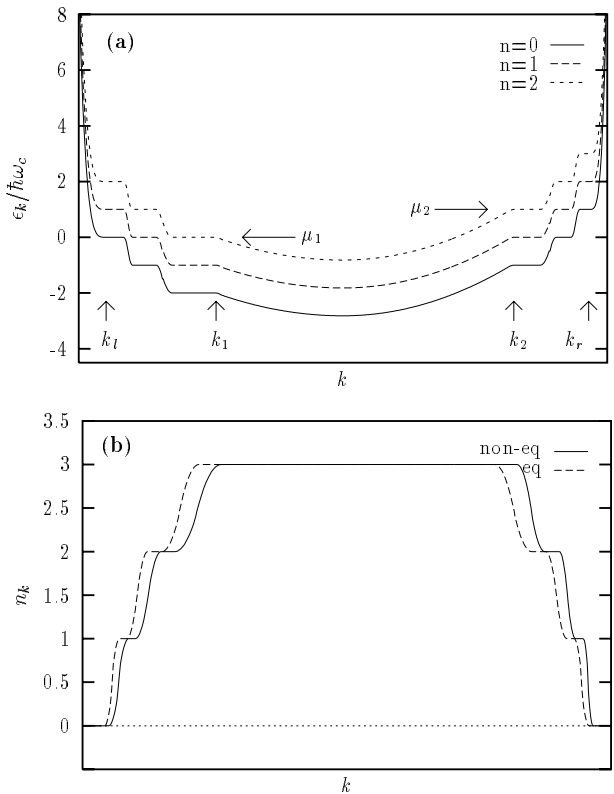

FIG. 1. (a) Energy for the non-equilibrium case. (b) Occupation number for equilibrium and non-equilibrium cases, note excess charge at the edges.

Consider now the case when current is injected into the sample. Under the conditions of QHE, longitudinal conductivity is negligible, and there is no current between the two edges. Therefore, for every (not too large) value of the injected current, there exists a many-particle steady state that carries the current. This non-equilibrium steady state can be described by certain charge and current distributions just like an equilibrium state. The only difference is that in the steady state the net current in the transverse direction is different from zero. As a consequence of this, Landau levels are filled up to different energies at the left and the right edges. As the non-equilibrium current is carried by the states in the incompressible regions, the total current flowing in the sample can be obtained by summing over the contributions due to the electrons in fully occupied states. The current carried by the one-particle state with wave vector $k$ can be written in terms of the group velocity $v_k$ and the density $\rho_k$ of electrons with this wave vector:

$$j_n(k) = e\rho_k v_k = \frac{e}{\hbar}\frac{\partial \epsilon_{n,k}}{\partial k}\int dx \mid \Psi_{n,k}(x,y)\mid^2 = \frac{e}{\hbar L_y}\frac{\partial \epsilon_{n,k}}{\partial k}. \quad (1)$$

Adding up all such contributions, one gets

$$I = \frac{e}{\hbar L_y}\sum_{n=1}^{N}\int_{k_l}^{k_r}dk\frac{L_y}{2\pi}\frac{\partial \epsilon_{n,k}}{\partial k} = \frac{e}{h}\sum_{n=1}^{N}\left(\epsilon_{n,k_r} - \epsilon_{n,k_l}\right). \quad (2)$$

Here $k_l$ ($k_r$) are the wave vectors of the last occupied state of the $n$th Landau level at the left (right) edge, respectively.

The Hall conductivity is $\sigma_{xy} = I/V_H$, where $V_H$ is the Hall voltage measured in the four-terminal resistance experiments. This voltage is determined by the amount of current carried by each Landau level just before the voltage contacts, as well as by the transmission and reflection coefficients of these contacts. It is possible to show [3] that the Hall conductivity will be quantized *only* if, upon leaving the voltage contacts, all Landau levels are filled up to the same energy: $\epsilon_{n,k_l} = \mu_1$, $\epsilon_{n,k_r} = \mu_2$ for all $n$. In this case, the voltmeter measures the difference between the chemical potentials at the two edges: $V_H = (\mu_2 - \mu_1)/e$. Then, according to Eq. (2), $\sigma_{xy} = Ne^2/h$. A corresponding picture of the occupation numbers and energy levels in the steady state is shown on Fig. 1.

Thus, in the QH regime each Landau level starts filling when the energy drops to $\mu_1$ (or $\mu_2$) at the left (right) edge. This means that *all* partially occupied states on one edge have the same energy ($\mu_1$ or $\mu_2$). Therefore, chemical potential starts dropping only after filling of the innermost compressible strip is completed (in other words, after the filling factor reaches its bulk value): $\mu$ changes only in the region $k_1 < k < k_2$ (Fig. 1a). Note that the existence of the wide incompressible bulk region is necessary but not sufficient: for the QHE to be observed, edge channels have to be completely equilibrated [3,14]. In the following by 'QH regime', we mean the situation when all edge channels are in perfect equilibrium with one another. We will later return to the point of *how* they actually equilibrate.

Here we address the main question of the letter and show that complete equilibration between the edge channels leads to a very specific spatial distribution of the Hall current. As discussed above, even in equilibrium, before injecting any current, there is an orbital current flowing along the edges. One needs to find out how the *injected extra current* is distributed in the sample; in particular, whether it flows along the edge channels where the gradient of the confining potential is large, or it runs through the bulk — the central region of constant occupation. Using Eq. (2), one finds, for instance, that the net left-edge current carried by the $n$th Landau level, i.e. current carried by the states with $k_l < k < k_1$ (see Fig. 1), is equal to

$$I_{n,edge} = \frac{e}{h}\left(\mu_1 - \epsilon_{n,k_1}\right). \quad (3)$$

This current includes both equilibrium and non-equilibrium portions. In order to separate non-equilibrium part, we first notice that the difference between the energies of two states with different $n$'s but the same $k$, is a multiple of $\hbar\omega_c$: $\epsilon_{n_1,k} - \epsilon_{n_2,k} = (n_1 - n_2)\hbar\omega_c$. If $N$ Landau levels are filled in the bulk of the sample, the energy of the state on the $N$th level with $k = k_1$ is equal to $\mu_1$. Therefore

$$\epsilon_{n,k_1} = \epsilon_{N,k_1} - (N-n)\hbar\omega_c = \mu_1 - (N-n)\hbar\omega_c. \quad (4)$$

Then, according to Eq. (3), the edge current carried by the $n$th Landau level is proportional to the energy difference between the states on the $N$th and $n$th Landau levels in the last compressible region:

$$I_{n,edge} = \frac{e}{h}(N-n)\hbar\omega_c. \quad (5)$$

This difference, however, cannot change by any noticeable amount when the current is injected into the sample. Therefore, the right-hand side of Eq. (5) is practically the same both in the equilibrium state and in the state with injected current. This leads us to the conclusion that in the QH regime the states in the edge channels carry the same amount of current as they did before the current was injected. Thus, only a negligible portion of the non-equilibrium current, associated with very small changes of the cyclotron frequency, flows in the edge channels. The rest of the current, or better to say, *almost the entire injected current is carried by the states in the bulk* ($k_1 < k < k_2$). As expected,

$$I_{bulk} = \frac{e}{h}\sum_{n=1}^{N}(\epsilon_{n,k_2} - \epsilon_{n,k_1}) = \frac{Ne}{h}(\mu_2 - \mu_1), \quad (6)$$

which is exactly the amount of the injected current.

Thus, QH regime is characterized by the absence of the Hall current in the edge channels and by the drop of the entire Hall voltage in the bulk of the sample. This is achieved by a redistribution of charge at the edges [6,7]: the widths of the compressible strips increase at the edge with lower chemical potential and decrease at the other edge (Fig. 1b). The electric field created by differently charged edges makes the bulk electrons move in the transverse direction.

Our conclusion on the spatial distribution of the injected current does not contradict the results of the experiments [10,11]. Both incomplete equilibration and non-local conductivity were observed not in the QH regime but rather in the situation when the Hall conductivity was not quantized and edge currents existed. Also, as stressed in [14], edge channel theory does not actually imply that the entire current flows at the edge. In the linear response regime this description leads to the correct results without specifying any particular current distribution.

Finally, we return to the question of how the equilibration between edge channels occurs. There are two ways for this: either due to tunneling between those channels, or due to the external 'mixer', namely the contact. Experiments with non-ideal contacts [10] — contacts that only allow some channels to interact with the contact — have shown that the first mechanism is not very effective, especially for the innermost channel, which is separated from the previous one by a relatively wide incompressible strip. One concludes, therefore, that edge channels must equilibrate *at the contacts* themselves for QHE to be observed. This is achieved due to inelastic scattering at the contacts.

Decades of industrial research have perfected 'ohmic' contacts between a metal lead and a semiconductor device, with very high transmission in both directions. We will refer to them as *ideal contacts*. One must consider the radically different environments on each side of the contact. On the device side one has a very clean 2D system, while on the lead side one has a dirty 3D metal with an enormous number of occupied and available states at the Fermi energy. It follows that one can consider the contacts to act as reservoirs of electrons for the device: all electrons that leave the device loose their memory upon entrance to the contact and all electrons that come from the contact and enter the device do so at the Fermi energy of the contact. The presence of the strong magnetic field further blocks backscattering in the semiconductor so that the transmission coefficient from the semiconductor side to the metallic side can be regarded to be essentially one. It is clear that after an ideal contact all channels must be equilibrated, at the same chemical potential as the contact itself (Fig. 2). Thus, the contacts play the crucial role of 'channel equilibrator'.

*Non-ideal contacts* are best realized by electrostatically pinching the 2DEG near the contact itself (Fig. 2).

This way one creates barriers in a very clean region and electron behavior can be simply modeled by transmission/reflection coefficients that depend on the Landau level index. The metal lead is still an ideal contact but is connected to and in equilibrium with a subsystem connected itself to the rest of the Hall device through these electrostatic barriers. Note how the absence of backscattering in the barrier region influences the way electrons are transmitted/reflected.

between different Landau levels required for establishing the QH regime occurs only at the contacts. These conclusions will hold even if there are FQH strips at the edges of the IQH liquid, since they are between regions with the same chemical potential.

We are grateful to Jung Hoon Han for numerous stimulating discussions. We are also grateful to I. Ruzin for sending us parts of his unpublished paper. This work was supported in part by the NSF Grant No. DMR-9220733.

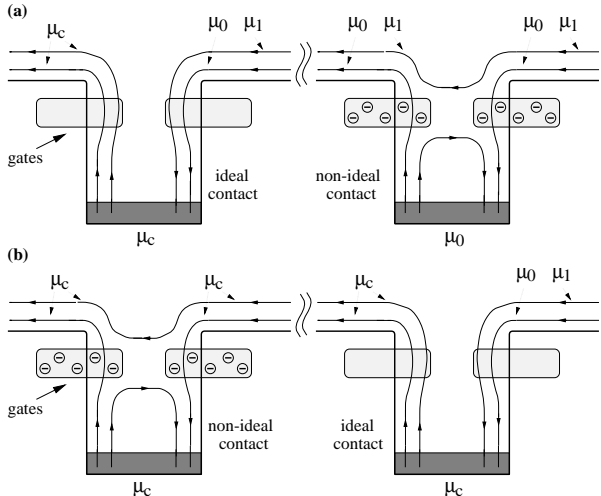

FIG. 2. Ideal and non-ideal contacts. (a) If a non-ideal voltage contact follows a non-ideal contact at the current source, resistance measured with this contact will not be quantized, while simultaneous measurement with the following ideal voltage contact will lead to a quantized resistance. (b) If the ideal contact is the first one on the path of electrons all edge channels are equilibrated after passing through it. Both measurement with the ideal contact and with the non-ideal contact following the ideal one lead to a quantized resistance.

One could test the equilibration at the contacts by carrying out six-terminal resistance measurement such that the contact at the current source and the adjacent voltage contact are non-ideal while the other voltage contact on the same edge is ideal (Fig. 2a). Hall conductivity measured with the non-ideal contact will not be quantized, while simultaneous measurement with the ideal contact will lead to the Hall quantization. On the other hand, when the direction of the injected current or of the magnetic field is reversed, both measurements should result in the quantized conductivity (Fig. 2b): the ideal contact is the first on the path of the electrons, and after passing through it all edge channels are equilibrated. Measurement with any kind of contact in the region after the ideal contact will lead to a quantized conductivity.

In conclusion, we showed that a specific feature of the QH regime is that almost the entire injected current is carried by states in the bulk of the sample. The role of the process of measurement is crucial, as equilibration

[1] K. von Klitzing, G. Dorda, and M. Pepper, Phys. Rev. Lett. **45** 449 (1980).
[2] B. I. Halperin, Phys. Rev. B **25**, 2185 (1982).
[3] M. Büttiker, Phys. Rev. B **38**, 9375 (1988).
[4] H. Aoki and T. Ando, Solid State Commun. **38**, 1079 (1981).
[5] J. Avron and R. Seiler, Phys. Rev. Lett. **54**, 259 (1985).
[6] D. J. Thouless, Phys. Rev. Lett. **71**, 1879 (1993).
[7] C. Wexler and D. J. Thouless, Phys. Rev. B **49**, 4815 (1994).
[8] T. Ando, Physica B **201**, 331 (1994).
[9] I. Ruzin, unpublished.
[10] B. J. van Wees et al., Phys. Rev. B **39**, 8066 (1989); B. W. Alphenaar, P. L. McEuen, R. G. Wheeler, and R. N. Sacks, Phys. Rev. Lett. **64**, 677 (1990); S. Komiyama, H. Hirai, S. Sasa, and F. Fujii, Solid State Commun. **73**, 91 (1990).
[11] P. L. McEuen et al., Phys. Rev. Lett. **64**, 2062 (1990); J. K. Wang and V. J. Goldman, Phys. Rev. B **45** 13479 (1992).
[12] N. Q. Balaban, U. Meirav, H. Shtrikman, and Y. Levinson, Phys. Rev. Lett. **71**, 1443 (1993).
[13] P. F. Fontein et al., Phys. Rev. B **43**, 12090 (1991), and references therein.
[14] C. W. J. Beenakker and H. van Houten, in *Solid State Physics*, edited by H. Ehrenreich and D. Turnbull (Academic, New York, 1991), Vol. 44; M. Büttiker, in *Semiconductors and Semimetals*, edited by M. Reed (Academic, New York, 1992), Vol. 35.
[15] C. W. J. Beenakker, Phys. Rev. Lett. **64**, 216 (1990); A. M. Chang, Solid State Commun. **74**, 871 (1990).
[16] D. B. Chklovskii, B. I. Shklovskii, and L. I. Glazman, Phys. Rev. B **46**, 4026 (1992).
[17] M. R. Geller and G. Vignale, Phys. Rev. B **50**, 11714 (1994).